\documentstyle[psfig]{europhys}

\def\etal{{\hbox{{\tenit\ et al.\/}\tenrm :\ }}}

\newif\ifboo \boofalse

\begin{document}
\euro{}{}{}{}
\Date{}
\shorttitle{A. PARISI \etal ROUGHNESS OF FRACTURE SURFACES}
\title{\Large Roughness of fracture surfaces}
\author{A. Parisi\inst{1,2}, G. Caldarelli\inst{1} and L. Pietronero\inst{1}}
\institute{
\inst{1} INFM - Unit\`a di Roma1 and Dip. Fisica, Universit\`a 
"La Sapienza", P.le A. Moro 2, 00185 - Roma, Italy\\
\inst{2} Department of Physics, University of Warwick, Coventry, CV4 7AL, UK}
\rec{}{}
\pacs{
\Pacs{62}{20.Mk}{}
\Pacs{05}{40.+j}{}
\Pacs{81}{40.Np}{}
}

\maketitle
\begin{abstract}
We study the roughness of fracture surfaces of three dimensional samples through 
numerical simulations of a model 
for quasi-static cracks known as Born Model. 
We find for the roughness exponent a value $\zeta \simeq 0.5$ 
measured for ``small length scales'' in microfracturing experiments. 
Our simulations confirm that at small 
length scales the fracture can be considered as quasi-static. 
The isotropy of the roughness exponent on the crack surface is also showed.
Finally, considering the crack front, we compute the roughness exponents of 
longitudinal and transverse fluctuations of the crack line
($\zeta_{\|} \sim \zeta_{\perp} \sim 0.5$). They result in agreement 
with experimental data, and support the possible application of the model 
of line depinning in the case of long-range interactions.
\end{abstract}

A large amount of studies has been devoted to the problem of material
strength and to the study of fractures in disordered media \cite{HerRou,CB}. 
In this paper we focus our attention on the self-affine properties of the 
fracture surface \cite{BLP,MPP}.
By self-affinity one means that the surface
coordinate $z$ in the direction perpendicular to the crack or fracture
(x-y) plane has the following scaling properties:
\begin{equation}
\begin{array}{rrr}
  z(\lambda x, y) & = & \lambda^{\zeta_x} z(x,y)  \\
  z(x, \lambda y) & = & \lambda^{\zeta_y} z(x,y)
\end{array}
\end{equation}
where the two $\zeta$ exponents are known as {\em roughness exponents}, and the $\hat{y}$
direction is chosen along the direction of propagation of the crack.
Even if these two directions seem to play a different role in the morphological
description of the fracture surface, experimental measurements showed that
these two directions have similar scaling properties for very different 
materials \cite{PKHRS}.
A unique roughness exponent is then
generally considered (therefore $\zeta_x = \zeta_y \equiv \zeta$),
and it has been claimed that it has a universal value 
of $\zeta=0.8$ \cite{BLP}. This behaviour has been confirmed 
for a large variety of experimental situations \cite{MHHR,SRB-94}.
However, more extended studies have also shown that in some experimental 
conditions, for fracture surfaces of metallic materials,
one deals with a different value of $\zeta=0.5$ \cite{MBSB}.
These two values of the roughness exponent are connected to the 
length scale at which the crack is examined. In particular, at small length 
scales one observes a roughness exponent $\zeta=0.5$, whereas at large 
length scales the larger value $\zeta=0.8$ is found. 
These results have recently been connected with the velocity of the crack 
front \cite{DNBC}, and interpreted as a quasi-static regime and a dynamic one: 
the value of $\zeta = 0.5$ should be expected in the quasi-static regime.  
This connection comes from the suggestion that the crack surface can be thought as 
the trace left by the crack front \cite{BBLP} so that the problem of the surface 
roughness can be mapped on the problem of the evolution of a line moving in
a random medium.
Erta\c{s} and Kardar \cite{EKa} introduce a couple of non-linear
Langevin equations to describe this evolution and their model can be usefully 
considered to evaluate the statistical properties of the evolution 
of a crack surface line. In this case, these equations describe longitudinal 
and transverse 
fluctuations with respect to the fracture plane containing the line velocity
and the pulling force \cite{B}. Values obtained are close to the roughness 
measured at large length scales.
At small length scales however, the crack front behaves as a moving line
undergoing a depinning transition, and in this case results from Erta\c{s}
and Kardar model \cite{EKb, MK-91, KZ-89}, should be revised \cite{B} 
considering long-range interactions, to give \cite{DNBC} 
$\zeta_{\|}$ and $\zeta_{\perp}$ equal to $0.5$ respectively 
for longitudinal and transverse fluctuations.
Another interesting model has been proposed by Roux et al. 
\cite{RF,HHR} where the fracture surface is expected to be a 
{\em minimal energy} surface. In two dimension this problem maps directly in the 
random directed polymer problem: 
the polymer with the minimum energy is the collection of the ``weakest'' 
monomers in the medium that form a directed path.
This corresponds to the surface crack of a fuse model and for brittle fractures 
gives a roughness exponent
$\zeta=2/3$ in 2 dimensions. Similar arguments hold for $d=3$ and are discussed for a scalar
model of fracture in \cite{RSAD}. In this case on has 
$\zeta = 0.42 \pm 0.01$, differently from experimental results \cite{RSAD}.

In this paper we present a numerical study of a fracture propagation for
a model of quasi-static fractures to show that in such a regime, the roughness exponent 
is in agreement with the one found at small length scales.
Comparisons with theoretical models are discussed at the end of this work.
To model the fracture propagation we use a mesoscopic model known as 
Born Model (BM), describing
the sample through a discrete collection of sites connected by springs 
\cite{YLS}.
The statistical properties of the two dimensional BM have been previously
considered in \cite{CCV,CDP}. In particular for $d=2$ \cite{CCG} a value of 
roughness of the fracture surface of $\zeta \simeq 0.64(3)$ in agreement with 
other measurements \cite{KHW} has been found.
In the BM the elastic energy of the sample under load is given by the energy 
of deformation for the springs connecting the sites of the sample. 
The elastic potential energy consists of two different terms, describing,
respectively, a central force and a non-central force contribution:
\begin{equation}
V=\frac{1}{2} \sum_{i,j} V_{i,j} = 
 \frac{1}{2} \sum_{i,j} \big\{ (\alpha - \beta) [({\bf u}_i-
{\bf u}_j) \cdot {\bf\hat{r}}_{i,j} ]^2
 +  \beta [{\bf u}_i-{\bf u}_j]^2 \big\}
\label{poten}
\end{equation}
where ${\bf u}_i$ is the displacement vector for site $i$,
${\bf \hat{r}}_{i,j}$ is the
unit vector between $i$ and $j$, $\alpha$ and $\beta$ are force
constants tuning the effects of central and non-central force contributions,
and the sum is over the nearest neighbour sites
connected by a non-broken spring.
By imposing the condition $\nabla_{{\bf u}} V=0$ one obtains a series of
equations for the fields ${\bf u}_i$. Solving them one obtains the 
equilibrium positions of the springs in the sample.
It has to be noticed that since one is interested in the equilibrium 
position one has to consider only the ratio of the two parameters
$\beta/\alpha$ (hereafter we will consider $\alpha=1$ and a varying $\beta$).
Simulations show that
varying $\beta$ between $1$ and $0$ corresponds to varying the Poisson 
coefficient between $0$ and $1/2$, as expected from the theory of elasticity.
At this point with a probability proportional to 
$\sqrt{V_{i,j}}$, which represents a generalized elongation,
one selects a spring to remove on the fracture boundary,
with the result of obtaining a connected fracture \cite{CCV, CCG};
by doing that the boundary
conditions of the system change and one has to compute a new equilibrium
position. Breaking a new bond after the complete relaxation of the lattice,
results in a slow velocity for the fracturing process, which mimics
a quasi-static process.
 
\begin{figure}
\protect
\centerline{
\psfig{file=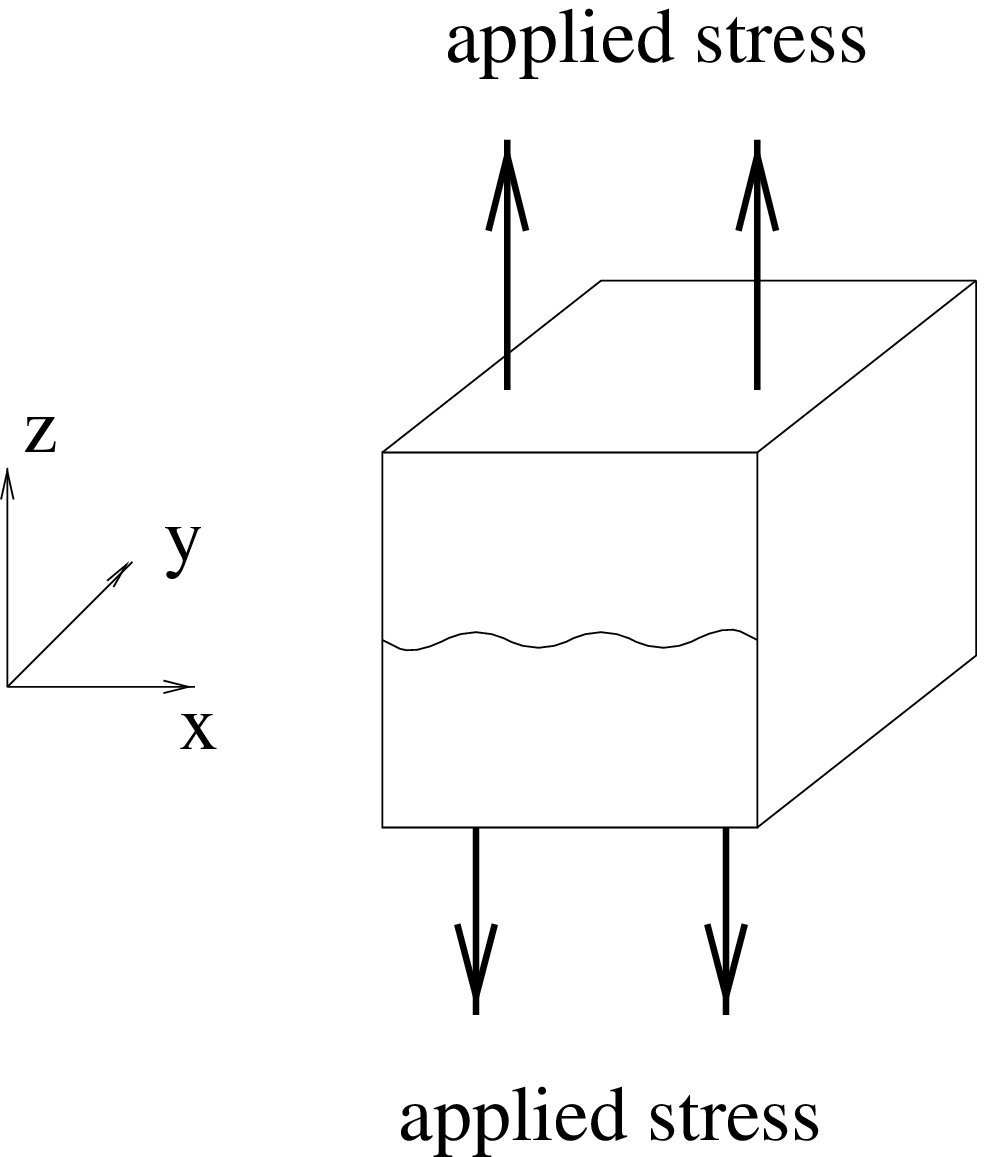,height=5cm,angle=0}
\hspace{1.5cm}
\psfig{file=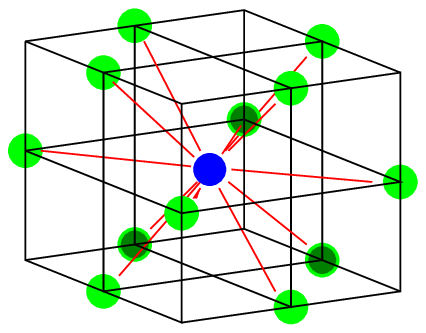,height=3cm,angle=0} 
}
\protect
\caption{On the left, the setup for the simulation, note the deterministic
starting notch.  On the right there is the FCC cell with the bonds used in 
the Born Model.}
\label{fig1} 
\end{figure}

The elastic springs can be arranged in different kinds of networks: however
in two dimensions one has to consider a triangular lattice, for a square 
lattice does not correctly describe the response of the system 
to the applied stress.
In three dimensions as well, one has to consider a network with the correct
response, which results in a more complex arrangement
with respect to the case of the simple cube.
In our case we chose a sample described by a Face Centered Cubic (FCC) 
structure and we applied a \mbox{mode I} loading on two opposite faces by
fixing their displacement field. 
We then applied periodic boundary conditions
on a second couple of opposite faces and on one of the 
two remaining faces we put a starting notch (see Fig.\ref{fig1}).

%
Starting from this setup we realized different simulations by stopping the
algorithm when the sample is divided in two parts. At this point we started
considering the surface of the fracture and we analyzed its statistical 
properties.
We performed simulations for different values of the $\beta$ parameter,
for 20 different samples of $32 \times 32 \times 16$ cells 
(each cell contains 4 sites for a total number of $2^{16}$ sites)
for each value of $\beta$, for which we obtained all the relevant results.
Further simulations on $40 \times 40 \times 20$ and 
$50 \times 50 \times 25$ FCC cells lattices were performed
to verify the generality of the results. 
Simulations lasted from a minimum of 18-hours of CPU time for each
$32 \times 32 \times 16$ cells lattice, up to
more than 180 hours for a $50 \times 50 \times 25$ cells lattice,
on a Digital alpha-station (500 MHz).

\begin{figure}[t]
\protect
\centerline{\psfig{file=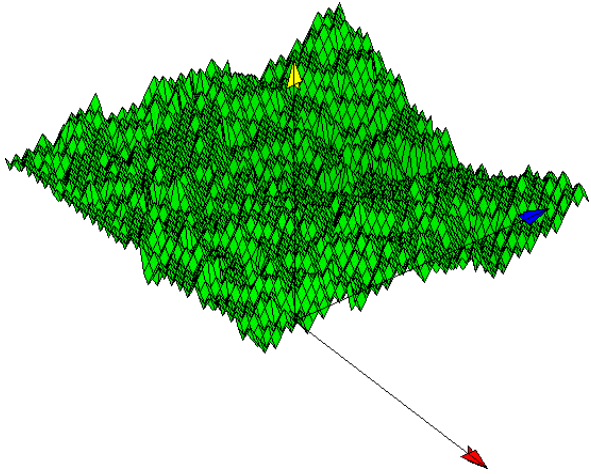,height=6.2cm,angle=0}} 
\caption{Fracture surface of $64 \times 64$ sites, obtained from a sample of 
$32 \times 32 \times 16$ FCC cells ($\alpha =1, \beta=1$).}
\label{fig2} 
\end{figure}

\begin{figure}[b]
\protect
\centerline{\psfig{file=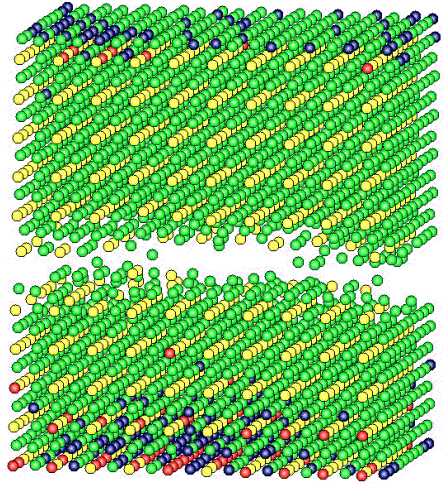,height=6.5cm,angle=0}}
\caption{An example of a just broken lattice of $10 \times 10 \times 10$ 
FCC cells (4000 sites: 4 site per cell). Here the fracture width has been 
enhanced for convenience, damaged sites are yellow and green the others are
blue and red. The double coloring (yellow/green and blue/red) is supplied to
better distinguish FCC cells.}
\label{fig3}
\end{figure}

An example of a final fracture surface is shown in Fig.\ref{fig2}, whereas
a typical broken sample is showed in Fig.\ref{fig3}: different colors show 
damaged (with at least one broken bond) and undamaged sites,
and the structure of the FCC lattice.

To compute the roughness exponent, we considered different cuts of the 
fracture surface, some of them along the $\hat{x}$ direction and some of 
them along the $\hat{y}$ direction which is the direction of propagation 
of the crack.
In principle we did not consider the fracture to be isotropic,
but we tried to recover the roughness exponents $\zeta_x$ and $\zeta_y$.
To measure the roughness of the surface, 
we followed the procedure
described in \cite{PKHRS},
by introducing the two spatial correlation functions
\begin{equation}
\begin{array}{rrr}
C_x(\rho)&=&\langle[z(x_i+\rho,y_i)-z(x_i,y_i)]^2\rangle \\
C_y(\rho)&=&\langle[z(x_i,y_i+\rho)-z(x_i,y_i)]^2\rangle
\end{array}
\end{equation}
where the average $\langle ... \rangle$ is taken over the different
$x$ and $y$ of the sites on the surface, and over different realization
of the surfaces.
Then we considered the power spectra $\tilde{C}_{x,y}(k)$ of the profile,
that is to say we studied the Fourier transform of the previous introduced 
correlation functions.
In this way the boundary effects are considered only in the large $k$ modes
\cite{BS}.
For self-affine profiles these power spectra are expected to scale as
\begin{equation}
\tilde{C}_{x,y}(k)\propto k^{-(1+2\zeta_{x,y})}
\end{equation}

\begin{figure}
\protect
\centerline{\psfig{file=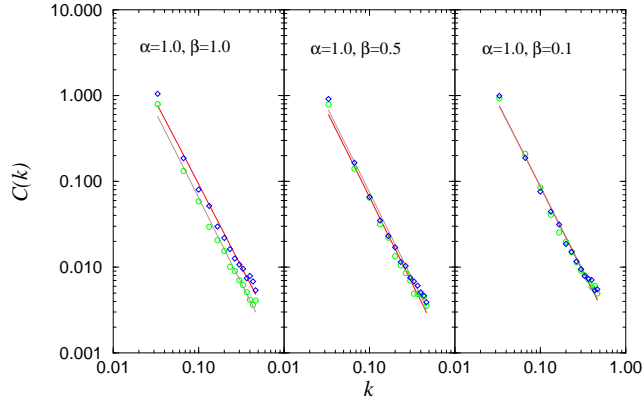,height=5.5cm,angle=0}}
\caption{Fourier transforms of the autocorrelation functions for the heights
for different values of $\beta$. All the fits have the same slope,
which corresponds to a value of $\zeta \sim 0.5$.}
\label{fig4}
\end{figure}

Fits for the Fourier transform are shown in Fig.\ref{fig4}.
Results show that the value of $\zeta_x$ is equal within the error bars 
to the value of $\zeta_y$. The two directions on the fracture surface 
show the same statistical properties as expected for a large variety of 
materials; the surface can then be described by a unique roughness exponent 
$\zeta$. Moreover, the value of $\zeta$ does not seem to depend on the value
of $\beta$, and is in complete agreement with the value expected
for fractures in a quasi-static regime \cite{DNBC}.
All the results are summarized in Tab.\ref{tab:tab1}.

\begin{table}
\centerline{
\begin{tabular}{|c|c|c|c|}
\hline
FCC cells & $\beta$  & $\zeta_x$ & $\zeta_y$ \\
\hline
$32 \times 32 \times 16$ & $1.0$ & $0.50\pm0.05$ & $0.49\pm0.05$ \\
                 & $0.5$  & $0.49\pm0.06$ & $0.46\pm0.06$ \\
                 & $0.1$  & $0.49\pm0.06$ & $0.47\pm0.05$ \\
                 & $0.01$ & $0.48\pm0.07$ & $0.47\pm0.06$ \\
\hline
$40 \times 40 \times 20$ & $1.0$  & $0.48\pm0.07$ & $0.47\pm0.06$ \\
\hline
$50 \times 50 \times 25$ & $0.5$  & $0.48\pm0.05$ & $0.48\pm0.05$ \\
\hline
\end{tabular}}
\protect 
\caption{Behavior of the two roughness exponents for different values
of the $\beta$ parameter, for FCC samples of different sizes.}
\label{tab:tab1}
\end{table}

\begin{figure}
\protect
\centerline{\psfig{file=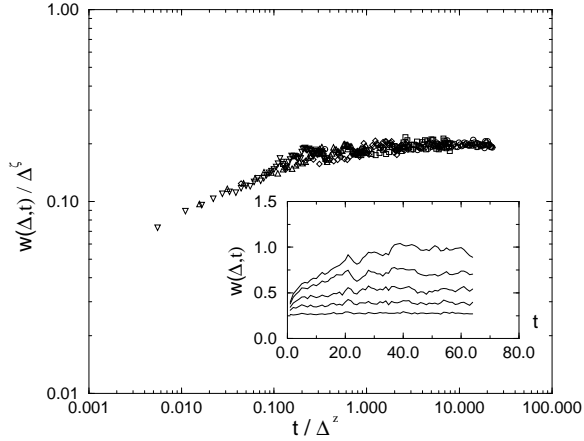,height=6cm,angle=0}}
\caption{Behaviour of the width functions: data are collapsed for $\zeta 
= 0.5$, and $z = 1.5$. Inset shows the width vs time for increasing sizes 
of $\Delta$.}
\label{fig5}
\end{figure}

To test such a measure (as suggested by Ref.\cite{SVR})
we also studied the scaling behaviour of the surface width
in direct space, along cuts perpendicular to the direction of the 
crack propagation:
the same
behaviour for $\zeta$ is recovered (see Fig.\ref{fig5}).  In this case results are quite striking, since a small deviation from the value of 0.5 leads to slightly displaced curves.  This observation allows us to be enough confident in these results even if they extend for about one decade.

An interesting analisys is the computation of the roughness of fluctuations of the crack front, compared with experiments and with the theory of line depinning by Erta\c{s} and Kardar.
It has to be noticed that the definition of the crack front has some sort
of ambiguity, because during the fracturing process more than one 
fracturing plane can develop, but just one among these will belong 
to the final fracture surface. Experimentally, this exponent can be found 
arresting the fracture during propagation, and injecting indian ink 
into the cracks under moderate vacuum. Samples are then dried and 
the process of fracturing is continued until complete separation
is reached \cite{B}.
From this point of view, one has to look at the border of the fracture 
belonging to what will be the final surface. Also, this corresponds 
to considering the fracture as the trace left from the crack front: 
this therefore belongs to the final surface.

Following this idea, we measured the roughness for the fluctuations of 
the crack front along the direction parallel to the line velocity 
($\zeta_{\|}$) and for those perpendicular to it ($\zeta_{\perp}$), 
during the crack evolution.
In the same way as the experimental one, we mark the sites reached through 
the fracturing process until a certain timestep. The process then continues 
up to complete separation. We then look at the marked part of the 
final surface, to recover the crack front.
Final results are from Fourier transform 
of an average at 
subsequent steps of the autocorrelation function in the steady state. 
In either cases we again obtained for both a value close to $0.5$ 
(see Tab.\ref{tab:tab2}), in good agreement with experiments. 
\begin{table}
\centerline{
\begin{tabular}{|c|c|c|c|}
\hline
FCC cells & $\beta$ & $\zeta_{\|}$ & $\zeta_{\perp}$ \\
\hline
$32 \times 32 \times 16$ & $1.0$ & $0.51\pm0.05$ & $0.49\pm0.06$ \\
                 & $0.5$  & $0.55\pm0.05$ & $0.48\pm0.06$ \\
                 & $0.1$  & $0.54\pm0.07$ & $0.50\pm0.07$ \\
                 & $0.01$ & $0.54\pm0.06$ & $0.51\pm0.07$ \\
\hline
$40 \times 40 \times 20$ & $1.0$ & $0.49\pm0.05$ & $0.49\pm0.08$ \\
\hline
$50 \times 50 \times 25$ & $0.5$ & $0.47\pm0.08$ & $0.46\pm0.09$ \\
\hline
\end{tabular}}
\protect
\caption{Measured values for the roughness of the crack front for 
different values of $\beta$. $\zeta_{\|}$ and $\zeta_{\perp}$ are
respectively the roughness for longitudinal and transverse fluctuations 
of the crack front. The large error bars for the $50 \times 50 \times 25$
samples are due to the limited number of sample examined, because of the 
lasting of simulations.}
\label{tab:tab2}
\end{table}

As regards the roughness exponent we can conclude from our simulations that its value is the one 
characteristic of the quasi-static evolution of cracks. 
This result is confirmed by quantitative analysis \cite{DHBC} and is to be
compared with results from the different approaches.
Our conclusions seem to be different from the conclusions of the directed 
polymers approach as presented with their scalar problem in \cite{RSAD}. 
The value of 
$\zeta = 0.42 \pm 0.01$ found with this model could
be related to the different physics of the fuse networks. The use of this 
model in fact, comes from the assumption that in two dimensions a fracture 
can be described by means of a scalar model. This is not stated in three
dimensions, where a description like the one that comes from the theory 
of elasticity can be obtained only through a vectorial model.
This could also explain the difference between our results, which agree with
experimental values, and those for the fuse network in \cite{BH-98}.

The analysis of the roughness of longitudinal and transverse fluctuations of the 
crack front, give results still in agreement with experiments and seems to 
confirm the mapping of the crack front to a line undergoing a pinning-depinning 
transition.
Our result is also to be compared with the one from \cite{REF-97}. 
In this paper in fact it is stated that an explanation of the roughness in 
terms of a quasi-static fracturing process seems unlikely. 
This seems to suggest a different conclusion, as in our simulations
elastic waves are cut-off through relaxation of the lattice after 
each bond-breaking. However, fluctuations are still enclosed in the 
stocastic process for the fracture.

In conclusion we presented numerical simulations of the fracture of a three
dimensional sample. Our result supports the idea that the fracture roughness
exponent is related to the different length scale  at which the sample is
analysed and then to the different dynamics of the crack.
In particular for  short length scales where the fracture can be identified
as quasi-static the roughness exponent is $\zeta=0.5$.
We also show that for elastic fractures one can expect isotropic behaviour
in the developing of the surface: our results show no dependance from 
the direction on crack surface.
As regards the crack front, our results agree with experimental measurements,
and support the mapping to line depinning in the case of long-range
interactions.

We whish to thank A.Petri and R.C.Ball for useful discussions. We also 
acknowledge support of EU contract N. ERBFMRXCT980183

\end{document}